\documentclass[prx,twocolumn,showpacs,A4]{revtex4}
\usepackage{amsmath}
\usepackage{dcolumn}
\usepackage{color}
\usepackage{bm}
\usepackage{youngtab}
\usepackage{graphicx}

\begin{document}

\title{Tensor renormalization of quantum many-body systems using projected
entangled simplex states}

\author{Z. Y. Xie$^1$\footnote{The first two authors contributed equally
to this work.}}
\author{J. Chen$^1$$^*$}
\author{J. F. Yu$^1$}
\author{X. Kong$^1$}
\author{B. Normand$^2$}
\author{T. Xiang$^1$}

\affiliation{$^1$Institute of Physics, Chinese Academy of Sciences,
P.O. Box 603, Beijing 100190, China}

\affiliation{$^2$Department of Physics, Renmin University of China,
Beijing 100872, China}

\begin{abstract}
We propose a new class of tensor-network states, which we name projected
entangled simplex states (PESS), for studying the ground-state properties of
quantum lattice models. These states extend the pair-correlation basis of
projected entangled pair states (PEPS) to a simplex. PESS are an exact
representation of the simplex solid states and provide an efficient trial
wave function that satisfies the area law of entanglement entropy. We
introduce a simple update method for evaluating the PESS wave function
based on imaginary-time evolution and the higher-order singular-value
decomposition of tensors. By applying this method to the spin-1/2
antiferromagnetic Heisenberg model on the kagome lattice, we obtain
accurate and systematic results for the ground-state energy, which
approach the lowest upper bounds yet estimated for this quantity.
\end{abstract}


\maketitle

\section{Introduction}

The theory of tensor-network states is evolving rapidly into an
interdisciplinary field involving condensed matter physics, quantum
information theory, renormalization group theory, and even quantum gravity.
From its initial proposals \cite{Niggemann1997, Sierra1998, Nishino2001},
through the development of representations and techniques \cite{PEPS, TEBD,
Vidal, LevinNave2007, Jiang2008, Xie2009, Zhao2010, Xie2012, Orus2013}, it
has become increasingly popular in the simulation of both classical models
\cite{Nishino2001, LevinNave2007, Xie2012, Chen2011} and strongly correlated
quantum systems \cite{Jiang2008, Xie2009, Zhao2010, TERG, Jordan2008,
LiWei2012}, providing deep insight into the physical properties of quantum
many-body states. In one dimension, the tensor-network state is known
as a matrix-product state (MPS) \cite{MPS}, and is also the wave function
generated by the density matrix renormalization group (DMRG) algorithm
\cite{DMRG}. A MPS may be viewed as a trial wave function arising from
virtual entangled pairs formed between two nearest-neighbor sites of a
lattice. Thus it yields a local description of quantum many-body states
based on their entanglement structure. A typical example of a MPS is the
$S = 1$ Affleck-Kennedy-Lieb-Tasaki (AKLT) state \cite{Affleck1987}, which
provides a prototype framework for understanding the physics of the Haldane
excitation gap in integer quantum spin chains.

Projected entangled pair states (PEPS) constitute a natural generalization
of MPS to two and higher dimensions \cite{PEPS}. This generalization, motivated
by two-dimensional AKLT states \cite{Affleck1987}, is obtained by distributing
virtual maximally entangled states between any two nearest-neighbor sites.
It leads to a faithful representation of the many-body wave function of
the ground state. Crucially, PEPS capture the boundary area law obeyed by
the entanglement entropy, which is believed to be the most important
ingredient causing quantum systems to behave differently from classical
ones \cite{Eisert2010}. It is precisely the existence of entanglement that
is responsible for such exotic phenomena as quantum phase transitions and
topological quantum order. Furthermore, PEPS allow a many-body ground-state
wave function, which contains exponentially many degrees of freedom, to be
calculated approximately but accurately on a polynomial time scale. In
particular, for a translationally invariant system, the understanding of the
whole wave function can be mapped to the problem of studying the properties of
just a single, or a small number of, local tensor(s).

Despite its strengths, the PEPS representation has two significant
disadvantages. It describes correctly the entanglement of adjacent basis
states, making it a good representation of AKLT-type states, and in
principle it can be used to represent all quantum states satisfying the
area law of entanglement. However, in practical calculations, the bond
dimension must be kept as small as possible to obtain sufficient accuracy
and efficiency, and this means that PEPS may not always provide a good
representation for the quantum states of some systems. As an example,
applying the PEPS algorithm on a triangular lattice is technically difficult
due to the high coordination number. A local tensor in PEPS on a triangular
lattice contains seven indices, six from the virtual bond degrees of freedom
and one from the physical degrees of freedom. Because the size of each tensor
scales as $D^6$, the bond dimension $D$ that can be handled practically by
current techniques is limited to a very small value (approximately 2$-$5).

The other disadvantage of PEPS concerns their application to frustrated
systems. They have been used to provide a very good variational ansatz for
the ground-state wave function of two-dimensional quantum spin models on
the square and honeycomb lattices \cite{Jiang2008,Xie2009,Zhao2010}. However,
for the antiferromagnetic Heisenberg model on the kagome lattice, we found
that the entanglement spectra of the local tensors for each one of the four
bonding directions are always doubly degenerate, due to the frustrated
lattice geometry, when $D > 3$. This causes a numerical instability that
is difficult to correct in the calculation of expectation values, and in
this case the PEPS ground-state energy does not converge with increasing
$D$. More generally, and as we discuss in detail below, it is difficult to
use PEPS to represent a quantum state in which the local correlation or
entanglement among all the basis states within a cluster (or simplex)
containing more than two lattice sites, for example the simplex solid
state proposed by Arovas \cite{Arovas2008}, becomes important.

In this work, we solve these problems by introducing a new class of
tensor-network states. We call these Projected Entangled Simplex States
(PESS), because they can be understood in terms of entangled simplex states
of virtual systems that are locally projected onto the physical basis states.
This class of states arises naturally as the exact representation of the
simplex solid states, but is of much broader use because, similar to PEPS,
any state can be represented by PESS if the virtual dimension is
sufficiently large. PESS extend pair correlations to simplex correlations
and hence constitute a natural generalization of the PEPS representation.

By the word ``simplex'' we refer to a cluster of lattice sites, which
constitute the basic unit, or ``building block,'' of a two- or
higher-dimensional lattice. As an example, a triangle is a building block
of the kagome lattice (Fig.~1) and can be taken as a simplex for this
lattice. However, one may also combine a number of simplices to form a
larger simplex; the choice of a simplex is not unique, but it should
reflect correctly the symmetry of the system. If a simplex contains $N$
lattice sites, we refer to the corresponding PESS as an $N$-PESS. If we
release the definition of the simplex and allow it to contain just two
neighboring sites, $N = 2$, the PESS are precisely the PEPS. Thus PESS
include PEPS as a subclass. As for PEPS, PESS are defined by introducing
a number of virtual basis states at each node of the lattice. In addition
to the local tensors, defined similarly to the PEPS framework for projecting
out the physical states from the virtual basis states at each node, the PESS
contain a new type of local tensor, which we call the ``entangled simplex
tensor.'' This tensor describes the correlation, or entanglement, of virtual
particles within the full simplex, and it is this feature that addresses
the frustration problem. An $N$-PESS with $N \ge 3$ is constructed as a
tensor-network product of these two types of local tensor. Examples of
this process are presented in Sec.~II.

Concerning the bond-dimension problem of PEPS, we provide a brief
example using the kagome lattice. The order (number of tensor indices)
of the local tensors in a PEPS representation is five and the size of the
local tensor is $dD^4$, where $d$ is the dimension of the physical basis
states. For PESS, as we will illustrate in Sec.~II, both types of local
tensors have only three indices, their sizes being $dD^2$ for the regular
projection tensors and $D^3$ for the entangled simplex tensor. Thus in
practical calculations a significantly larger bond dimension may be studied
in the PESS representation than by PEPS. While this is a major advantage of
PESS, it does not mean a PESS representation is always more efficient than
a PEPS one. For AKLT states, PEPS remain the most efficient representation,
whereas for simplex solid states, PESS are undoubtedly the most efficient.

We close this introduction by noting that general insight into the structure
of a quantum wave function may be obtained from singular value decomposition
(SVD). In the DMRG procedure in one spatial dimension, Schmidt decomposition
of the wave function is a SVD, and the SVD spectrum is simply the square root
of the eigenvalues of the reduced density matrix. Indeed, at the formal level
any wave function generated by DMRG can be expressed as a projected
``maximally entangled pair'' state; in this sense, the PEPS description
is equivalent to a SVD and the physical content of a MPS or PEPS ansatz can
be understood more generally from the entanglement structure of the wave
function under SVD. However, the PEPS approach offers a means of constructing
the wave function using only the local entanglement structure, which greatly
simplifies the construction of the PESS representation in comparison with
a SVD approach. This said, SVD \cite{Jiang2008,Xie2009,Zhao2010} and
higher-order SVD (HOSVD) \cite{Xie2012} of tensors is fundamental in
constructing renormalization schemes for tensor-network representations
of systems in dimensions higher than one, and is the core of the methods
employed in Sec.~IV.

This paper is arranged as follows. In Sec.~\ref{sec:SSS}, in order to
elaborate the physics underlying the PESS, we introduce an SU(2) simplex
solid state of spin $S = 2$ and construct explicitly both its PESS
representation and the parent Hamiltonian. In Sec.~\ref{sec:trialWF} we
propose the PESS as a trial wavefuntion for the ground states of quantum
lattice models. We introduce in Sec.~IV a simple update approach for
evaluating the PESS wavefunction based on the HOSVD of tensors. By applying
this approach to the spin-1/2 Heisenberg model on the kagome lattice, we
obtain the ground-state energy as a function of the bond dimension $D$ for
simplices with $N = 3$, 5, and 9. Section \ref{sec:summary} contains a
summary and discussion.

\section{PESS representation of simplex solid states}
\label{sec:SSS}

The simplex solid state of SU(N) quantum antiferromagnets was introduced
by Arovas \cite{Arovas2008}. It extends the bond singlets of the AKLT
state to $S = 0$ states of $N$-site simplices, with $N \ge 3$. Each simplex
accommodates a virtual quantum singlet. As with the AKLT states, the simplex
solid states are extinguished by certain local projection operators. This
feature allows one to construct a many-body Hamiltonian for which the simplex
solid state is an exact ground state, usually with a gap to all low-energy
excitations.

\begin{figure}[t]
\begin{center}
\includegraphics[width=8cm]{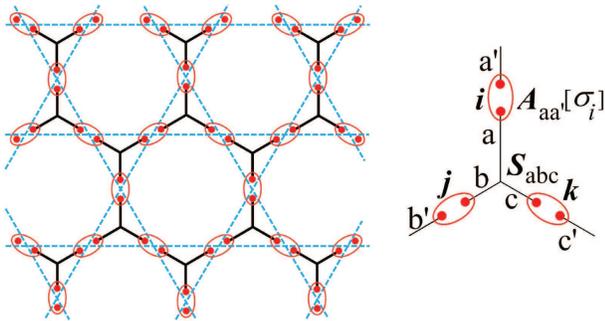}
\end{center}
\caption{(Color online) The spin-2 simplex solid state on the kagome
lattice (blue dashed lines). The entangled simplex tensors $S$ form a
honeycomb lattice (black solid lines) and the projection tensors $A$
are defined on the decorating sites of this honeycomb lattice.}
\label{fig:SSS-Kagome}
\end{figure}

The wave function of simplex solid states can be expressed as a
tensor-network state. This tensor-network state is the PESS, a result
we illustrate by constructing a simplex solid state and its PESS
representation for the $S = 2$ Heisenberg model on the kagome lattice.
The kagome geometry is a two-dimensional network of corner-sharing
triangles, each forming a three-site simplex. As shown in
Fig.~\ref{fig:SSS-Kagome}, the simplices form a honeycomb lattice,
on which the kagome lattice is formed by the decorating sites.

\subsection{Spin-2 kagome lattice}

A physical $S = 2$ state can be regarded as a symmetric superposition of
two virtual $S = 1$ spins. On the kagome lattice, two neighboring triangles
(simplices) share a single site. As in an AKLT state, we can assign each of
the $S = 1$ spins to one of the simplices associated with this site. There
are then three $S = 1$ spins on each simplex triangle, and their product
contains a unique spin singlet state,
\begin{equation}
\underline{1}\otimes \underline{1}\otimes \underline{1}=\underline{0}\oplus
\left( 3\times \underline{1}\right) \oplus \left( 2\times \underline{2}%
\right) \oplus \underline{3}.
\end{equation}%
This allows us to define a virtual singlet on the simplex,
\begin{equation}
\left\vert 0,0\right\rangle =\frac{1}{\sqrt{6}}\sum_{s_{i}s_{j}s_{k}}%
\varepsilon _{s_{i}s_{j}s_{k}}\left\vert s_{i}\right\rangle \left\vert
s_{j}\right\rangle \left\vert s_{k}\right\rangle,
\end{equation}%
where $\left\vert s_{i} \right\rangle$ $\left( s_{i} = -1,0,1 \right)$ is a
basis state of the $S = 1$ spin at site $i$ and $\varepsilon _{ijk}$ is the
Levi-Civita antisymmetric tensor.

The many-body state with this singlet on each simplex is a simplex solid
state. Its wavefunction, illustrated in Fig.~\ref{fig:SSS-Kagome}, is a
PESS, which can be expressed as
\begin{eqnarray}
\left\vert \Psi \right\rangle &=&Tr\left( \dots S_{abc}A_{aa^{\prime }}
\left[ \sigma _{i}\right] A_{bb^{\prime }}\left[ \sigma _{j}\right]
A_{cc^{\prime }}\left[ \sigma _{k}\right] \dots \right)  \notag \\ & & \;\;\;\;\;\,
\left\vert \dots \sigma _{i}\sigma _{j}\sigma _{k} \dots \right\rangle ,
\label{eq:ssswf}
\end{eqnarray}%
where the trace is over all spin configurations and all bond indices.
$S_{abc}$ is the entangled simplex tensor defined on the simplex honeycomb
lattice. The physical basis states $\left\{ \sigma_{i},\sigma_{j}, \, ...
\right\}$ are defined on the decorating sites of the honeycomb lattice
$\left\{ i,j, \, ... \right\} $ (i.e.~on the kagome lattice sites). The
Roman letters $\left\{ a,b, \, ... \right\}$ denote the virtual bond states.
Because the virtual spins in each simplex triangle form a spin singlet,
$S_{ijk}$ in this case is simply an antisymmetric Levi-Civita tensor,
\begin{equation*}
S_{ijk} = \varepsilon_{ijk}.
\end{equation*}
$A_{i,i^{\prime }} \left[ \sigma _{1} \right]$ is a $3 \times 3$ matrix, which
maps two virtual $S = 1$ spins onto an $S = 2$ physical spin, and whose
components are given by the Clebsch-Gordan coefficients of the SU(2) Lie
algebra,
\begin{eqnarray*}
A_{11} \left[ 2 \right] & = & A_{33} \left[ -2 \right] \; = \; 1, \\
A_{12} \left[ 1 \right] & = & A_{21} \left[ 1 \right] \; = \; A_{23}
\left[ -1 \right] \; = \; A_{32} \left[ -1 \right] \; = \; \frac{1}{\sqrt{2}},
\\ A_{13} \left[ 0 \right] & = & A_{31} \left[ 0\right] \; = \; \frac{1}
{\sqrt{6}}, \\ A_{22} \left[ 0 \right] & = & \frac{2}{\sqrt{6}},
\end{eqnarray*}%
while all other matrix elements are zero.

\begin{center}
\begin{table}[b]
\caption{Coefficients for projection operators in Eq.~(\protect
\ref{eq:3SiteProject}). }
\label{tab:3SiteProject}%
\begin{tabular}{ccccccc}
\hline
\vspace{-3mm} &  &  &  &  &  &  \\
$n$ & $\quad $ & $P_{4,n }$ & $\quad $ & $P_{5,n }$ & $\quad $ & $P_{6,n }$
\\ \hline
\vspace{-2mm} &  &  &  &  &  &  \\
1 &  & $\displaystyle -\frac{ 9}{440}$ &  & $\displaystyle \frac{1}{360}$ &
& $\displaystyle -\frac{1}{5544}$ \\
\vspace{-2mm} &  &  &  &  &  &  \\
2 &  & $\displaystyle \frac{1017}{61600}$ &  & $\displaystyle -\frac{173}{%
75600}$ &  & $\displaystyle \frac{5}{33264}$ \\
\vspace{-2mm} &  &  &  &  &  &  \\
3 &  & $\displaystyle -\frac{23}{6160}$ &  & $\displaystyle \frac{197}{362880%
}$ &  & $\displaystyle -\frac{731}{19958400}$ \\
\vspace{-2mm} &  &  &  &  &  &  \\
4 &  & $\displaystyle \frac{39}{123200}$ &  & $\displaystyle -\frac{547}{%
10886400}$ &  & $\displaystyle \frac{61}{17107200}$ \\
\vspace{-2mm} &  &  &  &  &  &  \\
5 &  & $\displaystyle -\frac{23}{2217600}$ &  & $\displaystyle \frac{41}{%
21772800}$ &  & $\displaystyle -\frac{1}{6842880}$ \\
\vspace{-2mm} &  &  &  &  &  &  \\
6 &  & $\displaystyle \frac{1}{8870400}$ &  & $\displaystyle -\frac{1}{%
43545600}$ &  & $\displaystyle \frac{1}{479001600}$ \\
\vspace{-2mm} &  &  &  &  &  &  \\ \hline
\end{tabular}%
\end{table}
\end{center}

For this $S = 2$ PESS representation, the total spins have the following
possibilities on any given bond of the kagome lattice,
\begin{equation}
\underline{2}\otimes \underline{2}=\underline{0}\oplus \underline{1}\oplus
\underline{2}\oplus \underline{3}\oplus \underline{4}.
\end{equation}%
The fact that each bond belongs to a simplex means that it cannot be in the
fully symmetric $S = 4$ state. Thus this PESS is an exact ground state of
the Hamiltonian
\begin{equation}
H = \sum_{\left\langle ij \right\rangle} P_{4} \left( ij \right),
\label{eq:2SiteParent}
\end{equation}
where $P_{4} \left( ij \right)$ is a projection operator projecting the spin
states on any nearest-neighbor bond $\langle ij \rangle$ onto a state with
total spin $S = 4$. $P_{4} \left( ij \right)$ can be expressed using the
local spin operators as
\begin{equation*}
P_{4} \left( ij\right) = - \frac{1}{280} T_{ij} + \frac{3}{1120} T_{ij}^{2}
- \frac{1}{2016} T_{ij}^{3} + \frac{1}{40320} T_{ij}^{4} ,
\end{equation*}
where $T_{ij} = (\mathbf{S}_{i} + \mathbf{S}_{j})^2$. We note here that
the spin-2 AKLT state on the kagome lattice is also the ground state of
this Hamiltonian. In fact, it can be shown that the PESS wave function
for this system, defined by Eq.~(\ref{eq:ssswf}), is identical to the
AKLT state \cite{rtpc}. This is a very special property of the case we
have chosen for illustration; in the general case, there is no AKLT-type
representation for most simplex solid states.

In the PESS of Fig.~\ref{fig:SSS-Kagome}, half of the virtual spins at the
three vertices on any given simplex are quenched to zero. Thus the total spin
on a simplex cannot exceed $S = 3$. If we allow the system to have three-site
interactions within each simplex, then it is straightforward to show that the
above PESS is also the ground state of the Hamiltonian
\begin{equation}
H = \sum_{\alpha } \left( J_{4} P_{\alpha ,4} + J_{5} P_{\alpha ,5} + J_{6} P_{\alpha,6}
\right),
\label{eq:3SiteParent}
\end{equation}%
where $\alpha $ represents a simplex triangle, $J_{4}$, $J_{5}$, and $J_{6}$
are non-negative coupling constants, and $P_{\alpha,S}$ is the operator
projecting a state at each simplex triangle onto a state with total spin $S$.
Using the spin operators on the three vertices of the simplex, $\left(
\mathbf{S}_{\alpha,1}, \mathbf{S}_{\alpha,2}, \mathbf{S}_{\alpha 3} \right)$,
$P_{\alpha,s}$ can be expressed as
\begin{equation}
P_{\alpha,S} = \sum_{n=1}^{6} P_{S,n} \left( \mathbf{S}_{\alpha,1}
 + \mathbf{S}_{\alpha,2} + \mathbf{S}_{\alpha,3} \right)^{2n}
\label{eq:3SiteProject}
\end{equation}
where the coefficients $P_{S,n}$ are given in Table~\ref{tab:3SiteProject}.

\subsection{Generalizations to different spins and lattice geometries}

\begin{figure}[t]
\begin{center}
\includegraphics[width=5.5cm]{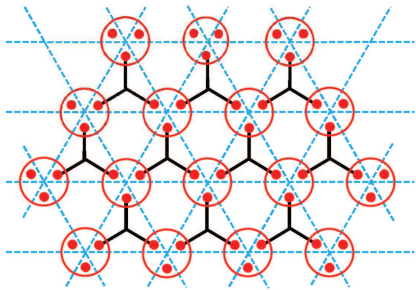}
\end{center}
\caption{(Color online) Schematic representation of the simplex solid state
on the triangular lattice (blue dashed lines).}
\label{fig:SSS-Triangle}
\end{figure}

\begin{figure}[t]
\begin{center}
\includegraphics[width=8cm]{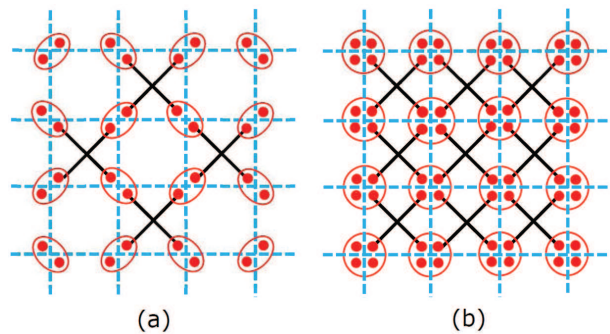}
\end{center}
\caption{(Color online) Schematic representation of two types of simplex
solid state on the square lattice (blue dashed lines). }
\label{fig:SSS-Square}
\end{figure}

The preceding discussion for the $S = 2$ simplex solid state can be
extended to systems of any higher spin, provided that a unique spin singlet
can be formed by the virtual spins in each simplex \cite{Arovas2008}. We
continue our illustration of the PESS representation by discussing briefly
its further generalization to describe simplex solids on different lattices,
choosing as examples the triangular (Fig.~\ref{fig:SSS-Triangle}) and square
(Fig.~\ref{fig:SSS-Square}) geometries.

For the simplex solid state on the triangular lattice shown in
Fig.~\ref{fig:SSS-Triangle}, the physical spin is formed by three virtual
spins. The simplex solid state is defined on a honeycomb lattice, which is
bipartite, with the simplex tensors on one of the sublattices and the
projection tensors on the other. If one assumes the virtual spin is still
in the spin-1 representation, then the physical spin will be in an $S = 3$
state. The simplex tensor is a $D = 3$ antisymmetric Levi-Civita tensor,
as for the kagome lattice. The projection tensor is now a four-indexed
quantity, with three virtual indices and one physical index. It maps three
virtual $S = 1$ states onto a fully symmetric $S = 3$ physical state. The
parent Hamiltonian for this PESS representation can be constructed in the
same way as for the kagome lattice. A parent Hamiltonian containing only
nearest-neighbor interaction terms is given by
\begin{equation}
H = \sum_{\left\langle ij\right\rangle} P_{6}\left( ij \right) ,
\end{equation}
where $P_{6} \left( ij \right)$ is the projection operator mapping the
two $S = 3$ states onto a state with total spin $S = 6$.

The definition of simplex solid states depends on the choice of simplex,
and on a given lattice is not unique. As an example of this, we show in
Fig.~\ref{fig:SSS-Square} that two kinds of simplex solid state can be
defined on the square lattice. If the lattice is taken as an edge-sharing
simplex lattice [Fig.~\ref{fig:SSS-Square}(b)], there are four virtual
particles on each lattice site and the simplex solid state so defined is
translationally invariant. The site projection is a five-indexed tensor.
If instead we take the square lattice as a vertex-sharing simplex lattice
[Fig.~\ref{fig:SSS-Square}(a)], then each site contains only two virtual
particles and the site projection is a three-indexed tensor. The order of
the projection tensors is also lower than the edge-sharing case. While this
simplex solid state is also translationally invariant, the lattice unit cell
is doubled.

The simplex solid state can also be considered in systems where the
generalized ``spin'' at each site has SU(N) symmetry, or obeys any other
Lie algebra. A general discussion of the SU(N) simplex solid states is
given in Ref.~[\onlinecite{Arovas2008}]. There is always a PESS
representation of simplex solid states and it is readily constructed from
the Clebsch-Gordon coefficients, or more generally from the decomposition
rules of the irreducible representations.

\section{PESS as a variational ansatz}
\label{sec:trialWF}

As for PEPS, it can be shown that PESS provide a good approximation for
the ground-state wave function, which satisfies the entanglement area law.
Thus PESS can be also regarded as a trial wave function for the ground
state of a quantum lattice model. To understand this statement clearly, we
take for illustration the spin-1/2 Heisenberg antiferromagnet on the kagome
lattice and demonstrate how to generate a PESS wavefunction by imaginary-time
evolution.

The Heisenberg model is defined on any lattice by
\begin{equation}
H = J \sum_{\left\langle ij \right\rangle} \left( S_{i}^{x} S_{j}^{x} + S_{i}^{y}
S_{j}^{y} + S_{i}^{z} S_{j}^{z} \right),
\end{equation}
where we take the simplest version in which $\left\langle ij \right\rangle$
denotes the summation only over all nearest neighbors. To perform the
imaginary-time evolution, we divide this Hamiltonian into a sum of three
terms,
\begin{equation}
H = H_{x} + H_{y} + H_{z},
\end{equation}
where
\begin{equation}
H_{\alpha} = J \sum_{\left\langle ij \right\rangle} S_{i}^{\alpha} S_{j}^{\alpha }
\end{equation}
with $\alpha = x, y, z$. All terms within $H_{\alpha }$ commute, but $H_{x}$,
$H_{y}$, and $H_{z}$ do not commute with each other. To evaluate the partition
function, we use the Trotter-Suzuki formula to decompose the evolution
opertor $e^{-\tau H}$ into a product of three terms,
\begin{equation}
e^{-\tau H} = e^{-\tau H_{x}} e^{-\tau H_{y}} e^{-\tau H_{z}} + {\cal O} \left( \tau^{2}
\right)
\end{equation}
for small $\tau$. In this approximation, the partition function can be
expressed as
\begin{eqnarray}
Z & = & Tr \, e^{-\beta H} \approx Tr \left( e^{-\tau H} \right)^{M}  \notag \\
 & \approx & Tr \left( e^{-\tau H_{x}} e^{-\tau H_{y}} e^{-\tau H_{z}} \right)^{M},
\label{epfa}
\end{eqnarray}
where $\beta = M \tau$.

We define a set of basis states specific to the spin-1/2 case,
\begin{equation}
\left\vert \sigma^{\alpha,n} \right\rangle = \left\{ \left\vert \sigma_{j}^{\alpha,n}
\right\rangle; j = 1, \dots, L \right\},
\end{equation}
where $\left\vert \sigma^{\alpha,0}
\right\rangle = \left\vert \sigma^{\alpha,M} \right\rangle$ and $L$ is the total
number of lattice sites. Here $\left\vert \sigma _{j}^{\alpha ,n}\right\rangle$
is the local basis state of $S_{j}^{\alpha }$,
\begin{equation}
S_{j}^{\alpha} \left\vert \sigma_{j}^{\alpha,n} \right\rangle =
\sigma_{j}^{\alpha,n} \left\vert \sigma_{j}^{\alpha,n} \right\rangle ,
\end{equation}
with eigenvalue $\sigma_{j}^{\alpha,n} = \pm 1$. By inserting these basis sets
into Eq.~(\ref{epfa}), we express the partition function in the form
\begin{eqnarray}
Z & \approx & \sum_{\left\{ \sigma^{x}, \sigma^{y} ,\sigma^{z} \right\}}
\prod\limits_{n=1}^{M} \left\langle \sigma^{x,n} \left\vert e^{-\tau H_{x}}
\right\vert \sigma ^{x,n} \right\rangle \left\langle \sigma^{x,n}| \sigma^{y,n}
\right\rangle \notag \\
& & \;\;\;\;\;\; \times \left\langle \sigma^{y,n} \left\vert
e^{-\tau H_{y}} \right\vert \sigma^{y,n} \right\rangle \left\langle \sigma^{y,n}
|\sigma^{z,n} \right\rangle  \notag \\
& & \;\;\;\;\;\; \times \left\langle \sigma^{z,n} \left\vert
e^{-\tau H_{z}} \right\vert \sigma^{z,n}\right\rangle \left\langle \sigma^{z,n}
|\sigma^{x,n-1} \right\rangle.
\label{eq:partition}
\end{eqnarray}
The basis sets $\left\vert \sigma^{y,n} \right\rangle$ and $\left\vert
\sigma^{x,n} \right\rangle$ are connected by the transformation matrix
$\left\langle \sigma^{x,n} | \sigma^{y,n} \right\rangle$, which is a product
of local transformation matrices at each site,
\begin{eqnarray}
A_{\sigma^{x,n},\sigma^{y,n}}^{x} & = & \left\langle \sigma^{x,n} | \sigma^{y,n}
\right\rangle = \prod\limits_{j} A_{j,\sigma^{x,n},\sigma^{y,n}}^{x},
\\
A_j^{x} & = & \frac{1}{2} \left( \begin{array}{cc}
1+i & 1-i \\
1-i & 1+i
\end{array} \right).
\end{eqnarray}
Similarly, one obtains for the other matrices
\begin{eqnarray}
A_{\sigma^{y,n},\sigma^{z,n}}^{y} & = & \prod\limits_{j} A_{j,\sigma^{y,n},\sigma^{z,n}}^{y}, \\
A_j^y & = & \frac{1}{\sqrt{2}} \left( \begin{array}{cc}
1 & -i \\
1 & i
\end{array} \right)
\end{eqnarray}
and
\begin{eqnarray}
A_{\sigma^{z,n},\sigma^{x,n-1}}^{z} & = & \prod\limits_{j} A_{j,\sigma^{z,n},\sigma^{x,n-1}}^{z},
\\ A_j^z & = & \frac{1}{\sqrt{2}} \left( \begin{array}{cc}
1 & 1 \\
1 & -1
\end{array} \right).
\end{eqnarray}

\begin{figure}[t]
\begin{center}
\includegraphics[width=8cm]{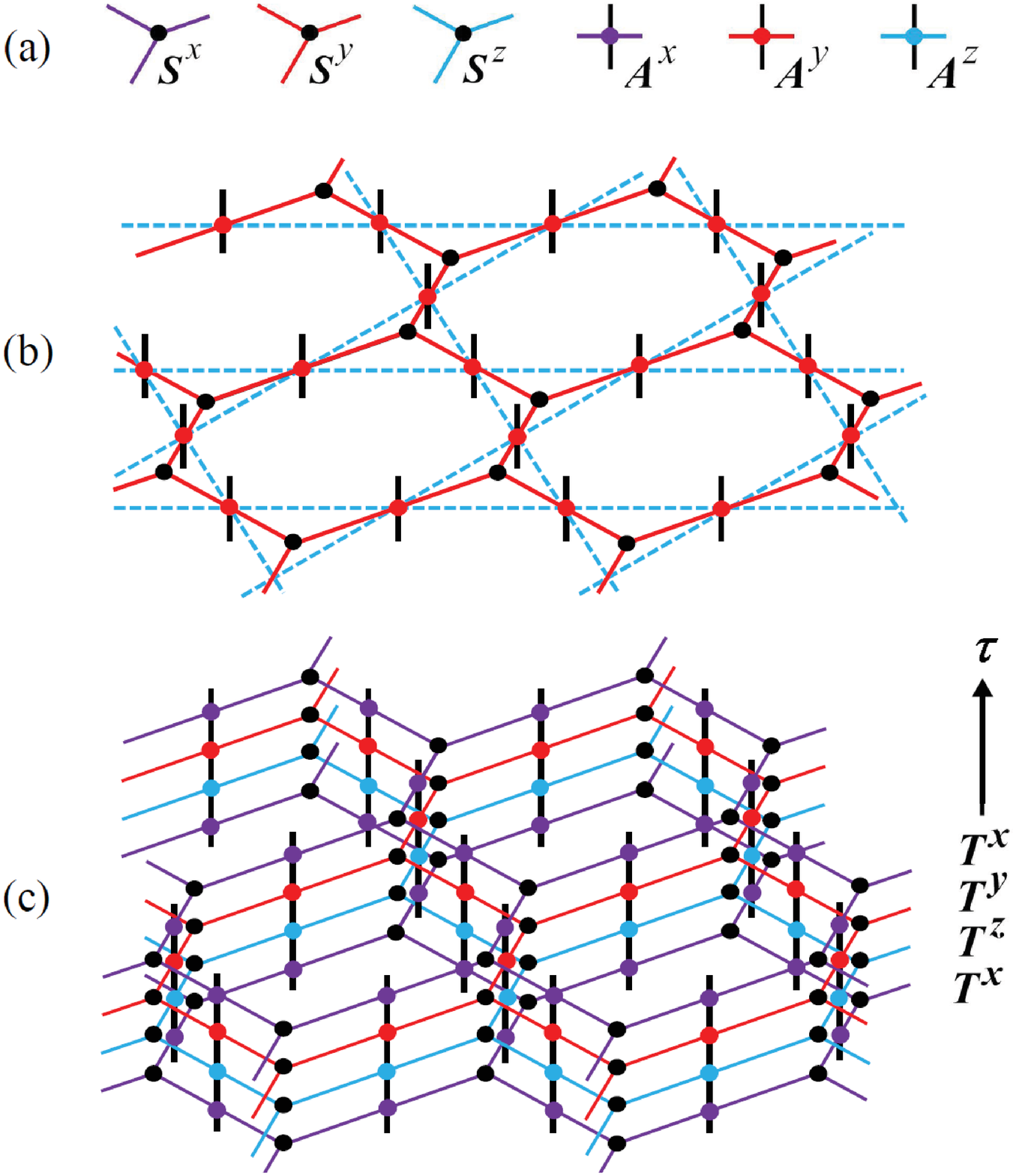}
\end{center}
\caption{(Color online) Graphical representation of simplex tensor operators
on the kagome lattice. (a) Simplex ($S$) and projection ($A$) operators. (b)
Tensor $T^z$ specified in Eq.~(\protect{\ref{eq:talpha}}). (c) One evolution
step of the simplex tensor operator $T= T^xT^yT^z$ describing
the partition function (\ref{eq:partition}).}
\label{fig:tensor-operator}
\end{figure}

In Eq.~(\ref{eq:partition}), $\left\langle \sigma^{\alpha,n} \left\vert
e^{-\tau H_{\alpha}} \right\vert \sigma^{\alpha,n} \right\rangle$ is the matrix
element of the classical Ising model $H_{\alpha }$, which is also the Boltzmann
weight of $H_{\alpha }$ for a given basis set $\left\vert \sigma^{\alpha,n}
\right\rangle $. As discussed in Ref.~[\onlinecite{Zhao2010}], this quantity
can be written as a product of local tensors,
\begin{equation}
\left\langle \sigma^{\alpha,n} \left\vert e^{-\tau H_{\alpha }} \right\vert
\sigma^{\alpha,n} \right\rangle = \prod\limits_{\nabla _{ijk}} S_{\sigma_{i}^{\alpha,n},
\sigma _{j}^{\alpha ,n},\sigma _{k}^{\alpha ,n}}^{\alpha },
\end{equation}
with
\begin{equation*}
S_{\sigma _{i},\sigma _{j},\sigma _{k}}^{\alpha }=\exp \left[ -\tau J\left(
\sigma _{i}\sigma _{j}+\sigma _{k}\sigma _{i}+\sigma _{j}\sigma _{k}\right) %
\right].
\end{equation*}
It is at this point where the lattice geometry enters, the symbol $\nabla$
indicating that the product is taken over all simplices (triangles) of the
kagome lattice.

Now the partition function becomes
\begin{equation}
Z\approx Tr \, T^{M},
\end{equation}
where $T = T^{x} T^{y} T^{z}$ is the tensor evolution operator and the matrix
elements of $T^{\alpha }$, given by
\begin{equation}
\left\langle \sigma^{\prime } \left\vert T^{\alpha } \right\vert \sigma
\right\rangle = \prod\limits_{\nabla _{ijk}} S_{\sigma_{i}^{\prime },\sigma_{j}^{\prime},
\sigma_{k}^{\prime }}^{\alpha } \prod\limits_{j} A_{\sigma_{j}^{\prime },\sigma _{j}}^{j} ,
\label{eq:talpha}
\end{equation}
contain both the entangled simplex and projection tensors. $T^{\alpha }$
defines a simplex tensor network operator on the decorated honeycomb lattice,
a graphical representation of which is shown in Fig.~\ref{fig:tensor-operator}.
Thus the partition function is expressed as a product of simplex tensor
network operators.

In the limit of zero temperature, $\beta \rightarrow \infty $, the partition
function (or the density matrix) is determined purely by the largest
eigenvalue and eigenvector of the evolution operator $T$. The largest
eigenvector may be found by the power method, starting from an arbitrary
initial wavefunction $\left\vert \Psi_{0} \right\rangle$, which is not
orthogonal to this eigenvector. Due to the simplex network structure of the
evolution operators $T^{\alpha }$, it is natural to assume that $\left\vert
\Psi_{0}\right\rangle$ is a PESS wave function. When $T$ is applied to
$\left\vert \Psi_{0} \right\rangle $, its PESS structure is retained, and
thus the ground-state wave function can indeed be expressed using PESS.
Of course, at each projection, or application of $T^{\alpha }$ to the wave
function, the bond dimension of the PESS is doubled. Thus in real
calculations the bond dimension must be truncated to find an approximate
PESS solution for the ground-state wave function.

\section{Simple update method for PESS calculations}
\label{sec:simple-update}

In principle, the PESS wave function can be determined by using the
variational approaches developed for PEPS. However, the bond dimensions
of PESS that can be treated with these techniques are generally very
small. An approximate but efficient means of determining the PEPS wave
function is the ``simple update'' method first proposed by Jiang
{\it et al.} \cite{Jiang2008}, which is in essence an entanglement
mean-field approach. It avoids a full calculation of the tensor environment
during the step where the wave function is updated by imaginary-time
evolution, which is usually the rate-limiting step in the calculation.
This procedure converts a global minimization problem into a local one,
yielding a fast algorithm that allows us to reach large values of the
bond dimension $D$. It is more effective for gapped systems and is almost
exact on Bethe lattices \cite{Vidal,LiWei2012} (a one-dimensional chain
can be regarded as the simplest Bethe lattice). However, the accuracy of
the results falls substantially when the system is close to a quantum critical
point, i.e. where correlations become long-ranged and full updating of the
environment tensor becomes essential \cite{LiWei2012}.

\begin{figure}[t]
\begin{center}
\includegraphics[width=8cm]{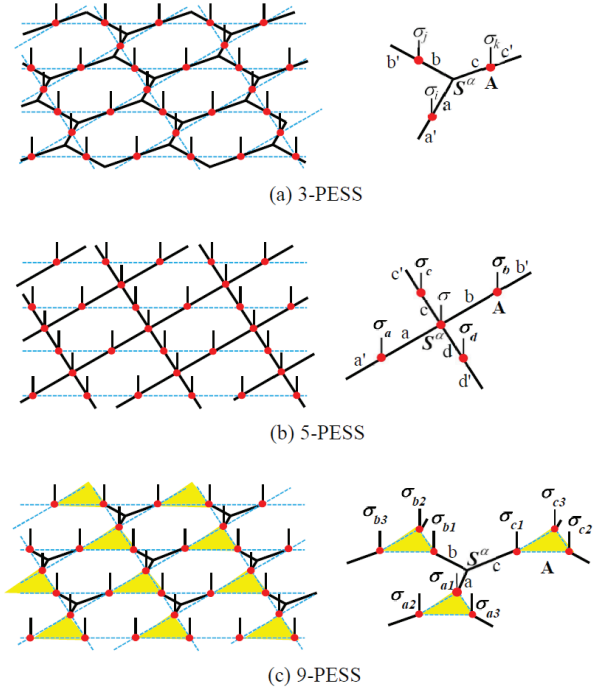}
\end{center}
\caption{(Color online) Graphical representations of a 3-PESS, 5-PESS, and
9-PESS on the kagome lattice (blue dashed lines). (a) The 3-PESS is defined
on the decorated honeycomb lattice. The vertical dangling bonds represent
the physical degrees of freedom $\{\sigma_i, \, ... \}$. $S^\alpha$ is
the entangled three-index simplex tensor and  $A$ is a three-index
tensor defined at each physical lattice site. (b) The 5-PESS is defined
on the decorated square lattice. The entangled simplex tensor $S^\alpha$ has
five indices, one of which represents the physical basis states at the nodes
of the square lattice while the other four represent the four virtual bond
states connecting to the neighboring decorated sites; it takes into account
all of the entanglement among these five spins. A tensor-network
ansatz with the same structure as this 5-PESS has been used in
Ref.~[\onlinecite{Corboz2012}] for studying the ground state of the SU(N)
model on the kagome lattice. (c) The 9-PESS is defined by taking the three
spins on each upward-oriented triangle as one effective site. The entangled
simplex tensor $S^\alpha$ has three indices and describes all of the
entanglement among the nine spins it connects.}
\label{fig:PESS-Kagome}
\end{figure}

Here we generalize the simple update method to study the PESS wave function,
by utilizing the HOSVD of tensors \cite{Xie2012,Lathauwer2000}. We again take
the kagome lattice to illustrate the method. Figure \ref{fig:PESS-Kagome}
shows graphical representations of the 3-PESS, the 5-PESS, and the 9-PESS,
which are the three simplest available PESS wave functions on the kagome
lattice. We stress again that each one takes into account all the correlations
(or entanglement) among the $N$ spins on the corresponding simplex, described
by the entangled simplex tensor $S$. In the limit of large $D$, this simplex
entanglement is treated rigorously. For simplicity, we describe only the
3-PESS in detail below. It is straightforward to extend the method to other
PESS representations and to different lattices.

We write the Hamiltonian in the form
\begin{equation}
H = H_{\triangle } + H_{\nabla },  \label{eq:HeisenbergKagome}
\end{equation}
where $H_{\triangle }$ and $H_{\nabla }$ are the Hamiltonians defined respectively
on all upward- and downward-oriented triangular simplices. As shown in
Fig.~\ref{fig:PESS-Kagome}, the 3-PESS is defined on a honeycomb lattice
formed by the simplex triangles. We assume the ground-state wave function
to be translationally invariant within each sublattice formed by the ``up''
or ``down'' triangle simplices, and hence that the simplex tensors are the
same on the same sublattice. The ground-state wave function may then be
expressed as
\begin{eqnarray}
\left\vert \Psi \right\rangle & = & Tr \left( \dots S_{a^{\prime}b^{\prime}c^{\prime}}
^{\alpha} A_{a^{\prime}a} \left[ \sigma_{i} \right] A_{b^{\prime }b} \left[ \sigma_{j}
\right] A_{c^{\prime }c} \left[ \sigma_{k}\right] \dots \right) \notag
\\ & & \;\;\;\;\;\; \left\vert \dots \sigma _{i} \sigma _{j} \sigma _{k} \dots
\right\rangle,
\label{eq:Psi}
\end{eqnarray}
where $\alpha$ represents the vertex coordinates of the simplex honeycomb
lattice and $S^{\alpha}$ is the corresponding entangled simplex tensor. As
in Sec.~II, the physical basis states $\left\{ \sigma_{i},\sigma _{j}, \,
... \right\}$ are defined on the kagome lattice sites $\left\{ i,j, \, ...
\right\}$ and the Roman letters $\{a,b, \, ... \}$ denote the virtual bond
states.

\subsection{HOSVD procedure}

As in Sec.~III, the ground-state wave function is determined by applying
the imaginary-time evolution operator $\exp \left( -\tau H \right)$ to an
arbitrary initial state $\left\vert \Psi_0 \right\rangle$, and in the limit
$\tau \rightarrow \infty$ the projected state $\exp \left( -\tau H \right)
\left\vert \Psi_0 \right\rangle $ will converge to the ground state. This
projection cannot be performed in a single step because the two terms in the
Hamiltonian~(\ref{eq:HeisenbergKagome}) do not commute with each other. To
carry out the projection, we take a small value for $\tau$ and apply the
evolution operator to $\left\vert \Psi_0 \right\rangle$ iteratively over
many steps. In the limit $\tau \rightarrow 0$, the evolution operator may
be decomposed approximately into the product of two terms by the
Trotter-Suzuki formula,
\begin{equation}
e^{-\tau H} = e^{-\tau H_{\triangle }} e^{-\tau H_{\nabla }} + {\cal O} \left( \tau
^{2} \right).
\end{equation}
Each projection is then performed in two steps, by applying $\exp(-\tau
H_{\triangle })$ and $\exp \left( -\tau H_{\nabla }\right)$ successively to
the wave function.

\begin{figure}[t]
\begin{center}
\includegraphics[width=7cm]{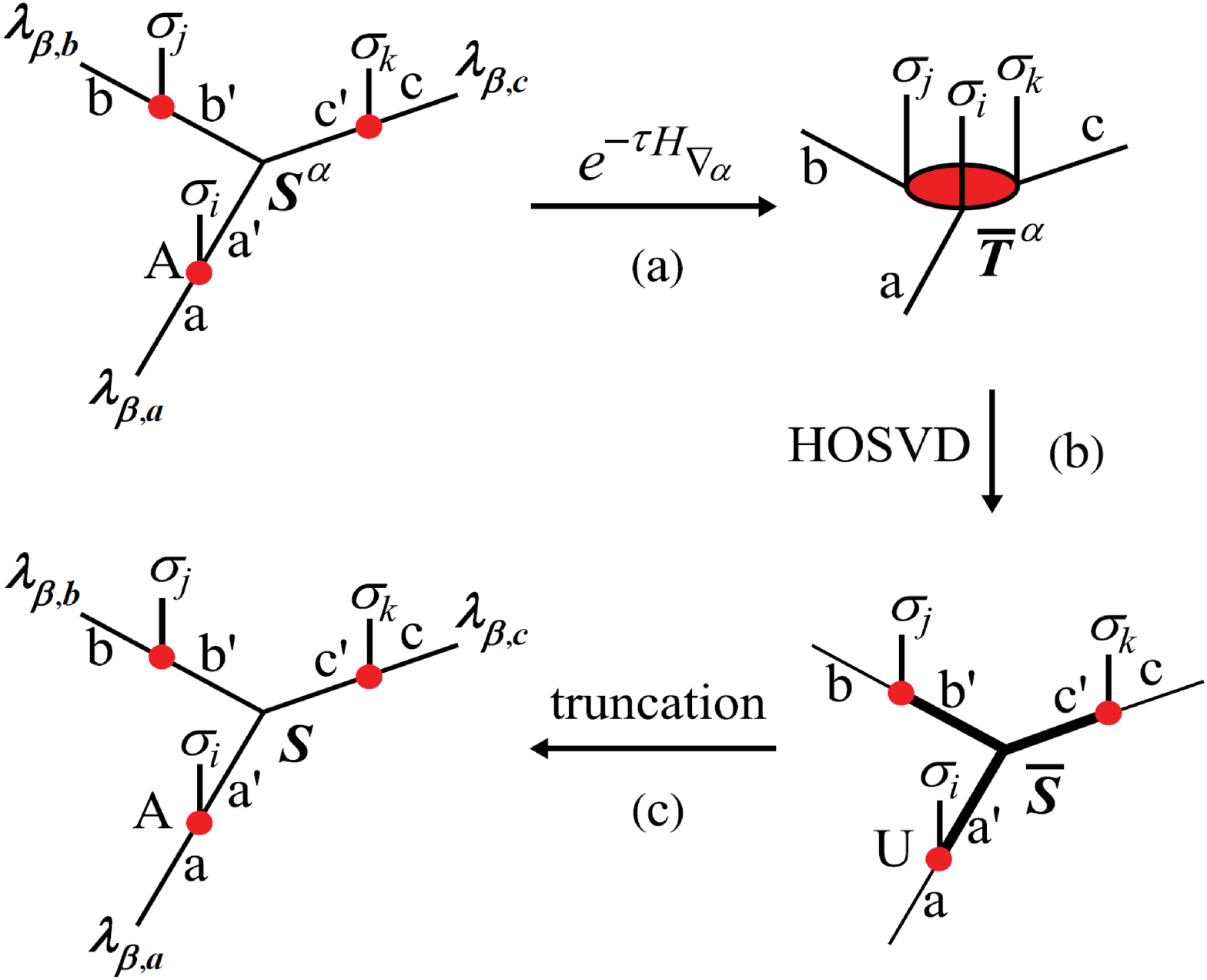}
\end{center}
\caption{(Color online) Flowchart for the simple update renormalization
scheme for the wave function using HOSVD. The environment contribution
around simplex $\alpha$ is described by a singular bond vector
$\protect\lambda_{\protect\beta}$ on each bond connected with the environment;
$\protect\lambda_{\protect\beta}$ is an approximate measure of the entanglement
on this bond. (a) $\exp (-\protect\tau H_{\protect\nabla\alpha})$ acts on the
tensors in a simplex to produce a new tensor $\overline{T}^{\protect\alpha}$
defined by Eq.~(\protect\ref{eq:Tnew}). (b) $\overline{T}$ is decomposed
by HOSVD [Eq.~(\protect\ref{eq:HOSVD})] into the product of a simplex tensor
$\overline{S}$ and three unitary matrices $U$. The dimensions of the thick
and thin black bonds are respectively $dD$ and $D$. (c) The thick bond
dimension is truncated from $dD$ to $D$, defining the renormalized $S$ and
$A$ tensors. }
\label{fig:FlowChart}
\end{figure}

We first consider the projection with $H_{\nabla }$. A schematic representation
of this procedure is shown in Fig.~\ref{fig:FlowChart}. Because all of the
separate terms in $H_{\nabla }$ commute with each other, the action of the
projection operator $\exp \left( -\tau H_{\nabla } \right)$ on a wave function
of the form specified by Eq.~(\ref{eq:Psi}) can be expressed as a product of
local evolution operators defined on each simplex (down triangle),
\begin{equation*}
e^{-\tau H_{\nabla }} \! \left\vert \Psi_0 \right\rangle = Tr ( \dots
T_{a\sigma_{i},b\sigma_{j},c\sigma_{k}}^{\alpha_\nabla} S_{ade}^{\beta_\triangle} \dots )
\left\vert \dots \sigma_{i} \sigma_{j} \sigma _{k} \dots \right\rangle ,
\end{equation*}
where $T_{a\sigma_{i},b\sigma_{j},c\sigma_{k}}^{\alpha_{\nabla}}$ is a $dD \times dD
\times dD$ tensor defined by
\begin{eqnarray}
T_{a\sigma_{i},b\sigma_{j},c\sigma_{k}}^{\alpha_\nabla} \!\! & = & \!\!
\sum_{\sigma_{i}^{\prime } \sigma_{j}^{\prime } \sigma_{k}^{\prime } a^{\prime} b^{\prime} c^{\prime }}
\!\!\!\!\!\! \left\langle \sigma_{i} \sigma_{j} \sigma_{k} \right\vert
e^{-\tau H_{\nabla\alpha}} \left\vert \sigma_{i}^{\prime} \sigma_{j}^{\prime}
\sigma_{k}^{\prime} \right\rangle \;\;\;\;\;\;\;\;\;\; \notag \\
& & \;\;\; \times S_{a^{\prime} b^{\prime} c^{\prime}}^{\alpha_{\nabla}} A_{a^{\prime }a}
\left[ \sigma_{i}^{\prime} \right] A_{b^{\prime}b} \left[ \sigma _{j}^{\prime }
\right] A_{c^{\prime }c}\left[ \sigma_{k}^{\prime} \right]
\label{eq:Tbar}
\end{eqnarray}
and $S_{ade}^{\beta_\triangle}$ represents the simplex tensors of the up triangles,
which are renormalized in the next step of the projection (below). In
Eq.~(\ref{eq:Tbar}), $H_{\nabla \alpha}$ is the Hamiltonian for the simplex
$\alpha$ and the local projection operator couples the simplex tensor
$S^{\alpha_{\nabla}}$ with the three neighboring $A$ tensors. For notational
simplicity, in the remainder of this section the superscript $\alpha$
refers to down triangles and $\beta$ to up triangles.

The next step is HOSVD, to decompose the tensor $T_{a\sigma_{i}, b\sigma_{j},
c\sigma_{k}}^{\alpha}$ into the product of a renormalized simplex tensor and
three renormalized projection ($A$) tensors. At this step one should also
include the renormalization effect of the environment tensors surrounding
$T^{\alpha}$ [Fig.~\ref{fig:FlowChart}]. Here we adopt an approximate
scheme to simulate the contribution of the environment tensors
\cite{Jiang2008} by introducing a positive singular bond vector
$\lambda_{\beta}$ (or $\lambda_{\alpha}$) of dimension $D$ on each bond
linking the $S^{\alpha}$ (or $S^{\beta}$) and $A$ tensors. This singular bond
vector may be determined iteratively by diagonalizing a density matrix $W$,
which is defined below, and it measures the entanglement between the
corresponding basis states on the two ends of the bond. This motivates
the definition of an environment-renormalized $T^{\alpha }$ tensor,
\begin{equation}
\overline{T}_{a\sigma_{i}, b\sigma_{j}, c\sigma_{k}}^{\alpha} = \lambda_{\beta,a}
\lambda_{\beta,b} \lambda_{\beta,c} T_{a\sigma_{i}, b\sigma_{j}, c\sigma_{k}}^{\alpha},
\label{eq:Tnew}
\end{equation}
where the three bonds of $T^{\alpha}$ are weighted by the corresponding
singular bond vectors. These additional bond vectors are included to
mimic the renormalization effect from the environment tensors in an
effective entanglement mean-field approach, which avoids the
(computationally expensive) full calculation of the tensor environment.

To truncate $\overline{T}^{\alpha}$ into a tensor of lower rank, we use a
HOSVD to decompose it according to
\begin{equation}
\overline{T}_{a\sigma_{i}, b\sigma_{j}, c\sigma_{k}}^{\alpha} = \sum_{a^{\prime}b^{\prime}
c^{\prime}} \overline{S}_{a^{\prime}b^{\prime}c^{\prime}}^{\alpha} U_{a^{\prime},a\sigma_{i}}
U_{b^{\prime},b\sigma_{j}} U_{c^{\prime},c\sigma_{k}} ,
\label{eq:HOSVD}
\end{equation}
where $\overline{S}^{\alpha}$ is the core tensor of $\overline{T}^{\alpha}$,
and satisfies two key properties for any given index. We illustrate these
using the second index $b$:

\noindent
(1) fully orthogonal:
\begin{equation*}
\left\langle \overline{S}_{:,b,:}^{\alpha} | \overline{S}_{:,b^{\prime},:}^{\alpha}
\right\rangle = 0,\text{ \ \ \ \ if } b \neq b^{\prime},
\end{equation*}
where $\left\langle \overline{S}_{:,b,:}^{\alpha}|\overline{S}_{:,b^{\prime},:}^{\alpha}
\right\rangle $ is the inner product of the two subtensors.

\noindent
(2) pseudodiagonal:
\begin{equation*}
\left\vert \overline{S}_{:,b,:}^{\alpha} \right\vert \geq \left\vert
\overline{S}_{:,b^{\prime },:}^{\alpha} \right\vert ,\text{ \ \ \ if }
b < b^{\prime},
\end{equation*}
where $\left\vert \overline{S}_{:,b,:}^{\alpha} \right\vert$ is the norm of
this subtensor, equal to the square root of the sum of squares of all
elements. These norms play a role similar to the singular values of the
matrix.

In Eq.~(\ref{eq:HOSVD}), $U$ is a unitary matrix of dimension $dD \times dD$,
determined by diagonalizing the density matrix
\begin{eqnarray*}
W_{a\sigma_{i},\underline{a}\underline{\sigma}_{i}} & = & \sum_{bc\sigma_{j}\sigma_{k}}
\overline{T}_{a\sigma_{i},b\sigma_{j},c\sigma_{k}}^{\alpha} \overline{T}_{\underline{a}
\underline{\sigma}_{i},b\sigma_{j},c\sigma_{k}}^{\alpha} \\
 & = & \;\, \sum_{a^{\prime}} U_{a^{\prime},a\sigma _{1}} \lambda_{\alpha,a^{\prime}}^{2}
U_{a^{\prime},\underline{a}\underline{\sigma}_{1}},
\end{eqnarray*}
where $\lambda_{\alpha,a^{\prime}}^{2}$ are the eigenvalues of $W$, which
measure the weights of the corresponding basis vectors $U_{a^{\prime}}$ in
$\overline{T}^{\alpha}$. With the aid of the $U$ matrices, we define
the renormalized $A$ tensor by
\begin{equation*}
A_{a^{\prime}a} \left[ \sigma \right] = U_{a^{\prime},a\sigma} \lambda_{\beta,a}^{-1},
\end{equation*}
where the dimension of the $a^{\prime }$ bond is truncated to $D$. Finally,
by keeping the first $D$ states for all three bond directions, we truncate
$\overline{S}^{\alpha}$ to a $D \times D \times D$ tensor $S^{\alpha}$.
This renormalized $S^{\alpha}$ tensor defines the new entangled simplex
tensor for its sublattice.

The projection with $\exp \left( -\tau H_{\triangle} \right)$ is performed in
the same way. By repeating this iteration procedure, an accurate ground-state
wave function is obtained after sufficiently many steps. The truncation error
in the tensors describing the ground-state wave function is reduced iteratively
throughout this renormalization procedure, and the iteration can be terminated
when the truncation error falls below a desired value.

\subsection{Ground-state energy for the spin-1/2 kagome antiferromagnet}

\begin{figure}[t]
\begin{center}
\includegraphics[width=10cm]{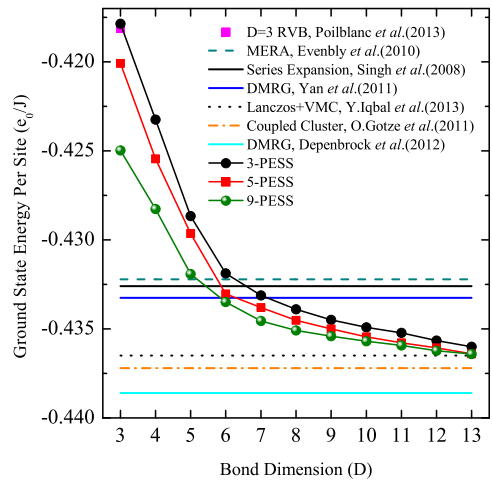}
\end{center}
\caption{(Color online) The dependence on bond dimension $D$ of the
ground-state energy of the $S = 1/2$ kagome Heisenberg antiferromagnet
obtained using 3-PESS, 5-PESS, and 9-PESS tensor-network representations.
The upper bounds on the ground-state energy obtained by a PEPS RVB ansatz
\cite{Poilblanc2012}, MERA \cite{Evenbly2010}, series-expansion methods
based on valence-bond crystal states \cite{Singh2007,Singh2008}, DMRG
\cite{Yan2011}, Lanczos exact diagonalization and VMC based on a gapless
Dirac spin-liquid state \cite{Iqbal2013}, and a high-order coupled-cluster
expansion \cite{Goetze2011}, as well as the DMRG result obtained by
extrapolation \cite{Depenbrock2012}, are shown for comparison. }
\label{fig:KagomeEnergy}
\end{figure}

We have applied the simple update scheme to the PESS representation of the
spin-1/2 Heisenberg antiferromagnet on the kagome lattice. The ground state
of this frustrated spin system has long been thought to be an ideal candidate
quantum spin liquid, a magnetic system with no spontaneous symmetry breaking
but showing specific topological order \cite{Balents2010}. This model has
been studied by approximate approaches for several decades \cite{Elser1989},
with many proposals for the nature of the ground state. Early numerical
calculations \cite{Marston1991} suggested that the ground state of the
model might be a valence-bond crystal, breaking the translational symmetry
of the kagome lattice, and this state has also been supported by
analytical arguments \cite{Nikolic2003}, by detailed cluster calculations
\cite{Singh2007,Singh2008}, by the multiscale entanglement renormalization
ansatz (MERA) \cite{Evenbly2010}, and by variational Monte Carlo (VMC)
studies \cite{Iqbal2011}. By contrast, different analytical arguments
\cite{Sachdev1992} and extensive DMRG studies \cite{Jiang2008sl,Yan2011,
Depenbrock2012} have all found the ground state to be a spin liquid with a
finite gap to triplet excitations; recent efforts to establish the topological
properties of this state \cite{Depenbrock2012,JWB2012} indicate that it is the
Z$_2$ spin liquid known from quantum dimer models. Other authors still have
suggested \cite{Ran2007,Iqbal2013} that the ground state is a gapless,
algebraic quantum spin liquid. We comment here that Poilblanc and coworkers
\cite{Poilblanc2012,Schuch2012,Poilblanc2013} have recently proposed a
PEPS-based trial wave function with resonating valence-bond (RVB) character
specifically to study the $Z_2$ spin-liquid phase. Their wave function, which
they found to work very well for this model, is actually a 3-PESS with $D = 3$.

Before presenting our results, we discuss the calculation of ground-state
expectation values using the PESS wave function. The calculation of the
wave function, as detailed in Secs.~III and IVA, is a fully variational
procedure and is subject to a truncation error that can be made arbitrarily
small by reducing $\tau$. To obtain an expectation value, we project the
wave function onto an MPS basis and calculate the required quantities
using the infinite time-evolving block-decimation method \cite{Vidal}.
While this procedure is not variational, the error in this part of the
calculation may be obtained by systematic variation of the bond dimension,
$D_{mps}$, of the MPS basis. When more than 60 basis states are retained
($D_{mps} > 60$), the truncation error due to the evaluation procedure is
less than 10$^{-4}$ for all of the 3-PESS and 9-PESS results shown in
Fig.~\ref{fig:KagomeEnergy}; however, it is somewhat higher for the 5-PESS,
where it varies up to a maximum of approximately $2 \times 10^{-3}$ for $D =
13$, even with $D_{mps} = 140$. We discuss this topic in further detail below.

Our result for the ground-state energy per site, $e_0$, of the kagome Heisenberg
antiferromagnet is shown in Fig.~\ref{fig:KagomeEnergy} as a function of
the bond dimension $D$, for the 3-PESS, 5-PESS, and 9-PESS representations
(Fig.~\ref{fig:PESS-Kagome}). As expected, the ground-state energy falls
with increasing $D$. In a gapped system, the ground state should converge
exponentially with $D$. However, the energies we obtain have not yet reached
the exponentially converged regime for any of the PESS representations, even
for $D = 13$. For this reason, we do not attempt an extrapolation to the
large-$D$ limit, because the results would be of limited meaning with the
available data and may be subject to significant errors. We stress that our
result is variational, hence representing an upper energy bound, and that
this bound can clearly be lowered quite significantly by further increasing
$D$. We remind the reader that our method is for a system infinite in size,
with truncation effected through $D$, and thus our results set an upper bound
for $e_0$ in the infinite two-dimensional limit. This new bound is the value
we obtain for the 9-PESS at $D = 13$, $e_0 = - 0.4364(1) J$.

In fact all three PESS values for the ground-state energy already lie
lower than the energies of the proposed valence-bond-crystal states
\cite{Singh2007,Singh2008,Evenbly2010}, and the best energy obtained by
contractor renormalization \cite{Capponi2013}, for $D = 7$. Larger values
of $D$ are required before the PESS values fall below the upper bound
obtained by DMRG in Ref.~\cite{Yan2011}. While the trend is clearly visible
in Fig.~\ref{fig:KagomeEnergy}, we have not yet been able to reach values
of $D$ sufficiently large that our calculated ground-state energy falls
below that obtained by the most sophisticated variational projector
quantum Monte Carlo calculations for the gapless spin-liquid scenario
($e_0 = - 0.4365 J$) \cite{Iqbal2013}, from the optimal extrapolated value
in the most detailed high-order coupled-cluster approach ($e_0 = - 0.4372 J$)
\cite{Goetze2011}, which favors a gapped spin liquid, or the approximate value
estimated recently from DMRG calculations by Depenbrock {\it et al.}, $e_0 =
 - 0.4386(5) J$ \cite{Depenbrock2012}. This last estimate may not be a true
upper bound for $e_0$ because it was obtained by an extrapolation of DMRG
results that continue to show a quite significant finite-size oscillation.

\begin{figure}[t]
\begin{center}
\includegraphics[width=8.5cm]{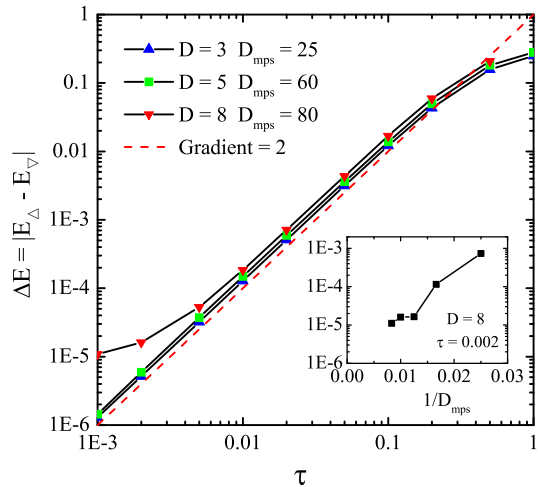}
\end{center}
\caption{(Color online)
Energy difference between up and down triangles in the 3-PESS, shown for
different bond dimensions $D$ as a function of Trotter step size $\tau$.
The dashed line shows the function $\tau^2$, indicating that the energy
difference vanishes in the limit $\tau \rightarrow 0$. For higher values
of $D$, calculations with higher values of $D_{mps}$ are required for full
convergence. Inset: the dependence of the energy difference on $D_{mps}$
shows a rapid convergence to values $D_{mps}$ of order $D^2$, followed by a
slow convergence dictated by $\tau$. }
\label{fig:PESS-sb}
\end{figure}

Regarding the qualitative properties of the ground state whose wave
function we have deduced, we make a further important comment concerning
its symmetry. The 3-PESS and 9-PESS break the symmetry between up and down
triangles, while the 5-PESS breaks the threefold rotational symmetry of the
kagome lattice. We have studied the energy differences induced in this way,
and illustrate their evolution with both Trotter step size $\tau$ and MPS
bond dimension $D_{mps}$ in Fig.~\ref{fig:PESS-sb} for the example of the
up-down asymmetry between triangles in the 3-PESS. This calculation also
illustrates the nature of the truncation error in the MPS calculation of
expectation values, where the effects of finite $\tau$, $D$, and $D_{mps}$
are interdependent. It is clear that the energy difference tends to zero,
restoring the symmetry of the ground-state wave function, as $\tau
\rightarrow 0$ at fixed $D$ and $D_{mps}$. This difference is in fact a
direct measure of the truncation error in the wave function, which is
fully controlled by $\tau$. The inset indicates that symmetry restoration
is also approached in the asymptotic limit of large $D_{mps}$, where again
it is limited by the value of $\tau$. Thus we can state with confidence
that the lattice symmetries are preserved in the true ground state. We
expect that physical quantities calculated from the PESS wave function,
including single-site magnetizations and single-bond spin correlation
functions, will show the same property of weak, symmetry-broken
differences tending to a uniform value in the appropriate limits.

We remark again that our ground-state wave function is obtained on the
basis of the simple update approximation. By adopting this procedure we
have essentially sacrificed a precise accounting for the effects of the
bond environment in exchange for the accuracy inherent in accessing larger
values of the tensor dimension $D$. Such an approach underestimates the
long-range correlation (entanglement) of the spins. To improve on this
result, and to calculate the correlation functions with the maximum accuracy
available within the PESS framework, one should perform a full-update
calculation taking complete account of the bond environment. We leave
this generally very time-consuming task for future study.

\section{Summary}
\label{sec:summary}

We have introduced the projected entangled simplex states (PESS) as a new
class of tensor-network states embodying the entanglement of particles
within a simplex. It is an exact tensor-network representation of the
simplex solid states first introduced by Arovas \cite{Arovas2008}. We have
demonstrated, using an SU(2) simplex solid state for $S = 2$ spins on the
kagome lattice, how to construct the PESS wave function and the parent
Hamiltonian. The discussion can be generalized to SU(N) or other groups
and to all lattice geometries.

PESS, together with projected entangled pair states (PEPS), form a
comprehensive representation of tensor-network states that satisfy the
area law of entanglement entropy \cite{Eisert2010}. They arise naturally
in the context of constructing trial wave functions for quantum systems
on two- or higher-dimensional lattices. For a wide variety of systems,
PESS provide an efficient representation of the exponential number of
coefficients by a small number of parameters describing the low-energy
physics of many-body quantum states arising from local interactions.
As for PEPS, PESS correlation functions are short-ranged, and so results
obtained with the PESS representation should converge exponentially with
increasing bond dimension $D$ for sufficiently large $D$ in a gapped system.
For a translationally invariant system, the PESS calculation is performed
directly on an infinite lattice, by-passing completely the errors
inherent in extrapolations from finite-size calculations.

PEPS and PESS are two types of trial wave function. In systems where the
correlation between pairs of neighboring sites are strongest, such as an
AKLT state, then PEPS are appropriate. If correlations among all the basis
states in a simplex or a larger cluster become important, then the PESS
representation is required. From our studies of the spin-1/2 kagome
Heisenberg antiferromagnet, the failure of PEPS to converge, contrasted
with the success of PESS, indicate that the effects of frustration in the
kagome geometry are well accounted for by the entangled simplex tensor
$S_{abc}$. An underlying reason for the success of the PESS wavefunction
on the kagome lattice may be that it is defined on the decorated honeycomb
lattice, which is geometrically unfrustrated. These observations suggest
that the problem of geometrical frustration in other lattices can be
similarly and approximately solved by finding a PESS representation whose
local tensors form an unfrustrated lattice.

PESS are also superior to PEPS in that the orders of the local tensors are
reduced in certain lattices. A particular example is the triangular lattice,
where in the PEPS representation the total number of tensor elements is
$dD^6$, while in a PESS representation (Fig.~\ref{fig:SSS-Triangle}) the
two tensors contain only $D^3$ and $dD^3$ elements. Still, a rigorous
evaluation of all local tensors in a PESS representation, including the
corresponding expectation values, requires a trace over all indices. This
is an exponentially hard problem and one not directly tractable for large
lattice systems, but approximate contraction schemes have been devised to
overcome this limitation. In the calculation of expectation values, there
is little difference between PEPS and PESS; the methods developed for
evaluating expectation values based on PEPS can be extended straightforwardly
to PESS.

To determine the PESS wave function, we have introduced a simple but
efficient update approach based on HOSVD. This is basically an entanglement
mean-field approach, which leads to a scalable variational method for finding
the local tensors. We have applied this method to the spin-1/2 Heisenberg
antiferromagnet on the kagome lattice and obtained an excellent estimate of
the ground-state energy, $e_0 = -0.4364(1)J$ (from the 9-PESS with $D = 13$).
This very promising result can be further improved by enlarging the order and
the bond dimension of the local tensors within the simple update scheme, or
more rigorously by a full update of the bond environment tensors. This latter
step will allow one to evaluate accurately the correlation functions and the
entanglement spectra. Efforts in this direction should help to make a
definitive identification of the topological phase in the ground state
of the kagome Heisenberg model.

The PESS representation can be readily extended to other lattices
and other models. It provides a significant advantage in studying the
ground-state properties of quantum lattice models on the triangular
(Fig.~\ref{fig:SSS-Triangle}), square (Fig.~\ref{fig:SSS-Square}), and
other lattices, because the order of the local tensors on these lattices
is much smaller than for the corresponding PEPS. In particular, we believe
that the PESS representation shown in Fig.~\ref{fig:SSS-Square}(b) offers
many advantages over PEPS for studying the $J_1$$-$$J_2$ antiferromagnetic
Heisenberg model on the square lattice \cite{Chandra1988, Melzi2000,
Jiang2012}. Finally, by proceeding as for the development of fermionic
PEPS \cite{Corboz2010, Kraus2010, Shi2009, Gu2010}, the PESS framework
can also be extended to include fermionic degrees of freedom.

\acknowledgments

We thank P. Corboz for valuable discussions, including those leading
us to correct Fig.~7, and W. Li, D. Poilblanc, and H. H. Tu for helpful
comments. This work was supported by the National Natural Science Foundation
of China (Grant Nos.~10934008, 10874215, and 11174365) and by the National
Basic Research Program of China (Grant Nos.~2012CB921704 and 2011CB309703).

\end{document}